\documentclass[twocolumn,secnumarabic,amssymb, nobibnotes, aps, prd]{revtex4-1}

\usepackage{graphicx}
\usepackage{slashed}
\usepackage{color}
\usepackage{soul}
\usepackage{bbm}
\usepackage{tensor}
\usepackage{dsfont}
\usepackage{hyperref}
\usepackage[all]{hypcap}
\hypersetup{colorlinks=true,
   linkcolor=blue,
    citecolor=red,
    filecolor=magenta,
    urlcolor=cyan}
\begin{document}

\title{Probing the curvature effects in graphene}%
\title{Pseudo-magnetic field in curved graphene}

\author{Pavel Castro-Villarreal}%
\email[e-mail: ]{pcastrov@unach.mx}
\affiliation{Facultad de Ciencias en F\'isica y Matem\'aticas, Universidad Aut\'onoma de Chiapas, Carretera Emiliano Zapata, Km. 8, Rancho San Francisco, C. P. 29050, Tuxtla Guti\'errez, Chiapas, M\'exico}
\author{R. Ruiz-S\'anchez}%
\affiliation{Universidad Polit\'ecnica de Chiapas, Carretera Tuxtla-Villaflores KM. 1+500, Las Brisas, 29150 Suchiapa, Chiapas, M\'exico}
\begin{abstract}
The general covariance of the Dirac equation is exploited in order to explore the curvature effects appearing in the electronic properties of graphene. Two physical situations are then considered: the weak curvature regime, with $\left|R\right|<1/L^2$, and the strong curvature regime, with $1/L^2\ll \left|R\right|<1/d^2$, where $R$ is the scalar curvature, $L$ is a typical size of a sample of graphene and $d$ is a typical size of a local domain where the curvature is pronounced. In the first scenario,
we found that the curvature transforms the conical nature of the dispersion relation due to a shift in the momentum space of the Dirac cone. In the second scenario, the curvature in the local domain affects the charge carriers in such a manner that bound states emerge; these states are declared to be  pseudo-Landau states because of the analogy with the well known Landau problem; here the curvature emulates the role of the magnetic field.  Seeking more tangible curvature effects we calculate e.g.  the electronic internal energy and heat capacity  of graphene in the small curvature regime and give an expresssion for the ground state energy in the strong curvature regime.\\

\noindent PACS number(s): 04.62.+y, 73.22 Pr, 71.70 Di
\end{abstract}
\maketitle

\section{Introduction}In the last decade, we have observed a paramount advance in the area of condensed matter physics: for the very first time an allotrope of carbon with the thickness of a single atom was synthesized. This allotrope has become popular under the name of graphene; in this material the carbon atoms are arranged in an hexagonal lattice \cite{Novolesov-Geim}. The graphene was imagined by P. R. Wallace in the fourties when he considered an isolated layer of carbon atoms as a model to understand the parallel and transverse conduction properties of graphite \cite{Ref-Wallace}. The synthesis of 2004 confirmed and found new exceptional pro\-per\-ties of graphene \cite{Castro-Neto, Vozmediano-Katnelson, Amorim-Cortijo-FJuan}. In particular, the experimental evidence confirmed that the charge carriers in  graphene can be described through quasi-particles that have the same behaviour as relativistic Dirac fermions \cite{Geim-Novolesoc}. This fact had been predicted theoretically when it was proved that the Dirac field theory in $2+1$ space-time dimensions emerges from Wallace's tight-binding model in the low energy regime (see \cite{Gordon-Semenoff} and \cite{divincenzo} for a related construction). It is noteworthy to mention that the Dirac structure emerges, in graphene as in the nowdays so called Dirac materials, as a consequence of specific symmetries \cite{diracmaterials}. The realization of graphene opens up new horizons in the production and the study of other two-dimensional materials like h-BN, germanene, and silicene, among others (see \cite{2dmaterials, analogo2d} for a review).

Experimentally, it has been shown that graphene presents corrugations even in free state \cite{Fasolino, Meyer, Geim}. Accor\-din\-gly, it has turned out that the Dirac equation in curved space is a natural model to study the electronic properties of the graphene sheet when corrugations and topological defects are taken into account \cite{Cortijo- Vozmediano}. In the last two decades, motivated by different problems in condensed matter physics, an intense activity has emerged in the study of the Dirac equation in curved space. This equation has been used to approach various problems like the spectrum of fullerenes \cite{Vozmediano-Guinea}, the problem of disclinations in icosahedral fullerenes \cite{kolesnikov},  the problem of impurities and topological defects in graphene \cite{Cortijo-Vozmediano}, the role of ripples in graphene \cite{JuanCortijoVozmediano}, the quantum Hall effect in topological insulators \cite{Lee}, the ultra-cold atoms on optical lattices \cite{BOADA}, the flexural phonons in curved graphene \cite{Kerner} and other gravitational analogues; problems that may have a physical realization in a bench-top experiment like wormholes in graphene \cite{ss},  BTZ black holes \cite{Gibbons}, and the Hawking-Unruh effect in graphene \cite{Iorio}, among others.  Some of the major changes correspond to the production of a gauge field similar to the electromagnetic field \cite{Vozmediano-Katnelson}. In particular the existence of pseudo-magnetic as a consequence of local deformations of graphene was first suggested theoretically \cite{guinea} and confirmed experimentally afterwards. In the scanning tunneling microscopy (STM) experimental results, Levy and coworkers showed the existence of pseudo-Landau le\-vels for fields greater than $300~{\rm T}$ in the corresponding triangular bubble-like structures created by the deformation \cite{levy}. 

The strain in graphene has resulted in a series of novel effects that are absent in flat graphene (see \cite{amorim} and references therein). Based on symmetry arguments of the fundamental lattice, in this last report,  all the interaction terms between the charge carriers and phonon on a sheet of deformed graphene that are linear in the strain tensor and its derivative were constructed in a very clean way. At a first glance \cite{FernandodeJuan-Sturla-Vozmediano}, this approach of introducing the strain in graphene is not compatible with the model based on the Dirac equation in curved space when the metric tensor is $g_{ij}=\delta_{ij}+2u_{ij}$, where $u_{ij}$ is the strain tensor, unless the curved surface is interpreted as the Dirac cone manifold \cite{Yang}. Here, we must emphasize that a model built using only symmetry arguments of the fundamental lattice may fail to account for the role of topological defects like the disclinations or pentagons formations due to the strain, which certainly break the hexagonal symmetry of the lattice \cite{Kolesnikov}.

In this paper, we explore the curvature effects on the electronic properties of a sheet of graphene based on the Dirac equation on curved spaces \cite{Cortijo- Vozmediano}. In order to study the effect of curvature we use the local density of states (LDOS), which is calculated using the Green's function approach \cite{Atland}. The general covariance of the corresponding Green equation and the Lichnnerowicz formula \cite{friedrich} are exploited in order to explore the curvature dependance of the Green's function. The calculation is done using the  coordinate frame provided by the Riemann normal coordinates (RNC) \cite{Eisenhart}. Namely, two physical situations of interest are considered:  a sheet of graphene almost flat with small corrugations and a sheet of graphene with  pronounced curvature in a local domain. These scenarios give rise to two different physical behaviors in LDOS; particularly, in the first situation we noticed that the curvature changes the conical nature of the dispersion relation resulting in a shift on the momentum space of the Dirac cone, while in the second situation we found that the geometric protuberance implies the existence of emergent Landau pseudo-states.  Here, as remarked below, we promote  the curvature, that is the Ricci (or equivalently the Gaussian curvature in this 2D case),  to be the fundamental physical quantity that characterizes the deformation of graphene sheet. Furthermore, we put forward the idea  that Ricci curvature is the actual physical quantity that gives origin to the pseudo-Landau levels  in analogy to the flat graphene case in presence of a uniform magnetic field. In this manner, the pseudo-magnetic field is proportional to the curvature itself.  Finally,  looking out for more tangible curvature effects,  we give expressions for the electronic internal energy and heat capacity  of graphene in the small curvature regime and the energy of the ground state in the strong curvature regime.
 
This paper is organized as follows. In section \ref{sec2}, the model for a curved graphene sheet is presented. The model is based on the Dirac equation defined on a space-time, $\mathcal{M}$ of dimension $2 + 1$. In section \ref{sec3}, the Green's function is computed using RNC in both physical regimes mentioned above. In section \ref{sec3dot5} we establish the analogy of the Dirac fermions in curved space with the Landau problem. In section \ref{sec4}, the curvature effects on the electronic properties of graphene are probed through the LDOS.  Finally, in section \ref{sec5}, we summarize the main conclusions and perspectives of this work.

\section{Curved Dirac model revisited}\label{sec2}

In this section, we introduce the simplest model to des\-cri\-be the electronic degrees of freedom in a curved sheet of graphene. To introduce the model, we start from the Dirac equation, 
\begin{eqnarray}
i\underline{\gamma}^\alpha\nabla_\alpha\psi=0,
\label{dirac_curvo}
\end{eqnarray}
defined on a $2+1$ curved space-time $\mathcal{M}$, where $\underline{\gamma}^{\alpha}\left(x\right)=\gamma^{A}e^{\alpha}_{A}\left(x\right)$. The $\gamma^{A}$ are the Dirac matrices that satisfy the Clifford algebra $\left\{\gamma^{A}, \gamma^{B}\right\}=2\eta^{AB}\mathbbm{1}_{2\times 2}$,  where $\eta_{AB}$ is the Minkowski metric tensor. 
The $\{e^{\alpha}_{A}\left(x\right)\}$ is a set of vierbeins attached to each coordinate patch of $\mathcal{M}$. The capital latin indices $A$ label  the Minkowski flat coordinates whereas the greek indices $\mu$  label the local curved coordinates.  In the Dirac equation $\nabla_{\alpha}=\partial_{\alpha}+\Omega_{\alpha}$ is the covariant derivative for the spinor representation of the Lorentz group $SO(2,1)$, where $\Omega_{\alpha}=\frac{1}{8}\omega^{AB}_{\alpha}\left[\gamma_{A}, \gamma_{B}\right]$ is the spin-connection and $\omega^{AB}_{\alpha}$ are elements of a $1$-form that satisfy the Maurer-Cartan equations \cite{Nakahara}. Thus $\Omega_{\alpha}$ and $e^{A}_{\alpha}\left(x\right)$ encode the geometrical content of the Dirac equation. 

The general space-time metric tensor can be written in terms of vierbeins, i.e. 
{\small $g_{\alpha\beta}=e^{A}_{\alpha}\left(x\right)e^{B}_{\beta}\left(x\right)\eta_{AB}$}. For simplicity, we consider a stationary space-time with curved space only,  $ds^2=-v^2_{F}dt^2+g_{ij}dx^{i}dx^{j}$ with $i,j=1,2$, where $v_{F}$ is the Fermi velocity. Now, we separate the time component from the spatial ones of the indices $A$ and $\mu$,  that is $\{0,a\}$ and $\{0,j\}$, respectively. Thus, suitable vierbeins for this metric are $e^{0}_{0}=1$, $e^{i}_{0}=e^{0}_{a}=0$ and $e^{i}_{a}\neq 0$, such that, $g_{ij}=e^{a}_{i}e^{b}_{j}\delta_{ab}$. For this metric, it is not difficult to show that 
$\omega^{0}_{j~b}=\omega^{a}_{0~b}=\omega^{a}_{j~0}=0$ and $\omega^{a}_{j~b}\neq 0$, with spatial indices $a,b,i,j$ and $k$. Therefore, the covariant derivative is reduced to
\begin{eqnarray}
\nabla_{0}&=&\partial_{0},\nonumber\\
\nabla_{j}&=&\partial_{j}+\frac{1}{8}\omega_{j}^{ab}\left[\gamma_{a},\gamma_{b}\right].
\end{eqnarray}
Next, the separation between temporal and spatial indices, in the Dirac equation, turns out to be a Schr$\ddot{\rm o}$dinger type of equation 
\begin{eqnarray}
i\hbar\partial_{0}\psi=\mathcal{H}_{K}\psi, 
\label{theDirac}
\end{eqnarray}
where $\mathcal{H}_{K}=-i\hbar v_{F}\gamma_{0}\underline{\gamma}^{j}\left(x\right)\nabla_{j}$. This is the same decomposition  performed in \cite{BOADA} for ultracold atom systems. Now, we have the ingredients  to write down a field-theoretic Hamiltonian for the curved sheet of graphene:
\begin{eqnarray}
\hat{H}=\int d^{2}x\sqrt{g}\Psi^{\dagger}\left(\begin{array}{cc}
\mathcal{H}_{K}&0\\
0&\mathcal{H}_{K^{\prime}}
\end{array}
\right)\Psi,
\label{model}
\end{eqnarray}
where $\mathcal{H}_{K^{\prime}}=(\mathcal{H}_{K})^{T}$ and $T$ is  matrix transposition. The letters $K$ and $K^{\prime}$ are refering to the two independent Dirac points.  A particular representation of the Dirac matrices is $\gamma^{0}=-i\sigma_{3}$, $\gamma^{a}\gamma^{0}=\sigma^{a}$, where $\sigma_{3}$ and $\sigma^{a}$, with $a=1,2$, are the standard Pauli matrices. In this representation, for flat graphene it is straightforward to show that $\mathcal{H}_{K}=-i\hbar v_{F}\sigma\cdot\partial$, i. e.  the low energy limit of the Wallace tight-binding model \cite{Castro-Neto}. Also, this Hamiltonian $\hat{H}$ is the same cosmological model posed by M. A. H. Vozmediano and coworkers  \cite{Cortijo- Vozmediano}.

\section{Green function for the curved sheet of graphene}\label{sec3}
Given fields $\Psi\left(x,t\right)$ that satisfy Eq. (\ref{theDirac}) through an standard Fourier transformation in time we get
 $\left(\epsilon-\mathcal{H}_{K}\right)\tilde{\Psi}\left(x,\epsilon\right)=0$, where $\tilde{\Psi}\left(x,\epsilon\right)$ is the time-Fourier transformed field.  A general field configuration is  obtained using the Green function, $\mathbb{G}\left({\bf x}, {\bf x}^{\prime}, z\right)$ of $z-\mathcal{H}_{K}$, where $\mathcal{H}_{K}=-i\hbar v_{F}\gamma_{0}\slashed{D}$ with $\slashed{D}=\underline{\gamma}^{j}\left(x\right)\nabla_{j}$ by the standard technique \cite{MorseAndFeshback}. Here, however, we will use this Green function in order to give an expression for the Local Density of States (LDOS) (see equation (\ref{LDOS})) of the curved sheet of graphene.

The general covariance of the Green equation is exploited to explore the curvature dependance of the Green function.  We shall use the local frame defined by the so-called Riemann normal coordinates (RNC) $y=x-x^{\prime}$, where $x^{\prime}$ is a fiducial point that can be chosen as the origin \cite{Eisenhart}. To this end, we revisit the calculation of the Green function in these coordinates. Let us first express $\mathbb{G}\left({\bf x}, {\bf x}^{\prime}, z\right)=\left(z+\mathcal{H}_{K}\right)G\left({\bf x}, {\bf x}^{\prime},z\right)$, where $G=G_{F}/\left(\hbar v_{F}\right)$. Consequently, $G_{F}$ satisfies the equation
\begin{eqnarray}
\hbar v_{F}\left(z^2_{F}-\mathcal{H}^2_{F}\right)G_{F}\left({\bf x}, {\bf x}^{\prime}, z_{F}\right)=\frac{1}{\sqrt{g}}\delta\left({\bf x}-{\bf x}^{\prime}\right)\mathbbm{1},
\end{eqnarray}
with the reduced quantities $z_{F}=z/\left(\hbar v_{F}\right)$ and $\mathcal{H}_{F}=\mathcal{H}_{K}/\left(\hbar v_{F}\right)$. 
Further simplification is gotten by taking the square of $\mathcal{H}_{F}$, since  the Clifford algebra implies  $\mathcal{H}^{2}_{F}=-\gamma_{0}\slashed{D}\gamma_{0}\slashed{D}=-\slashed{D}^2$, and using the Lichnerowicz formula \cite{friedrich} we obtain 
\begin{eqnarray}
\mathcal{H}^{2}_{F}=-\left(\nabla_{i}g^{ij}\nabla_{j}-\frac{1}{4}R~\mathbbm{1}\right),
\label{complete}
\end{eqnarray}
where $R$ is the scalar curvature. 

Following \cite{Parker}, we perform the change $\hbar v_{F}G_{F}(x,x^{\prime})=g^{-\frac{1}{4}}\bar{G}_{F}$, and additionally the factor $1/\sqrt{g}$ in the Dirac delta $\delta\left({\bf x}-{\bf x}^{\prime}\right)$ can be replaced by $g^{-1/4}$ in order to atain  further simplification \cite{deWitt}.   Now, the Green equation, after these changes, is cast into
\begin{eqnarray}
\left(z^2_{F}-\hat{h}^{2}\right)\bar{G}_{F}=\delta\left(y\right)\mathbbm{1}, 
\end{eqnarray}
where $\hat{h}^{2}=-g^{-\frac{1}{4}}\nabla_{i}g^{\frac{1}{2}}g^{ij}\nabla_{j}g^{-\frac{1}{4}}+\frac{1}{4}R$.  We can appreciate now  the analogy with a quantum mechanical problem with ``Hamiltonian''  $\hat{h}^{2}$. This operator can be further simplified as
\begin{eqnarray}
\hat{h}^{2}=-\nabla_{i}g^{ij}\nabla_{j}-g^{-\frac{1}{4}}\partial_{i}\left(g^{\frac{1}{2}}g^{ij}\partial_{j}\left(g^{-\frac{1}{4}}\right)\right)+\frac{1}{4}R.
\end{eqnarray}

Next, we use the fact that the metric tensor and spin connection (using RNC) can be written as a Taylor series with coefficients given in terms of the covariant derivative of the Riemann tensor \cite{Schubert}. The first terms of these series expansions are
\begin{eqnarray}
\label{metric}
g^{ij}=\delta^{ij}-\frac{1}{3}R\indices{^{i}_{k\ell}^{j}}y^{k}y^{\ell}+\cdots,\\
\Omega_{j}=\frac{1}{2}y^{k}R\indices{^{ab}_{kj}}\Sigma_{ab}+\cdots,
\label{spin}
\end{eqnarray}
where $\cdots$  are higher order terms $O(y^{n})$ with $n\geq 3$. After some calculations, this means that the ``Hamiltonian''  $\hat{h}^{2}$ can be written as
\begin{eqnarray}
\hat{h}^{2}=-\delta^{ij}\nabla_{i}\nabla_{j}+\frac{1}{3}R\indices{^{i}_{kl}^{j}}\partial_{i}\left(y^{k}y^{l}\partial_{j}\cdot\right)+\frac{1}{12}R+\cdots,\nonumber\\
\label{eqexact}
\end{eqnarray}
where curvature coefficients are evaluated already at the fiducial point $x^{\prime}$. This expression for $\hat{h}^{2}$ is exact but not so easy to manage because it involves an infinite number of terms.  Nevertheless,  Eq. (\ref{eqexact}) is our starting point to carry out approximations for $G_{F}\left({\bf x}, {\bf x}^{\prime}, z\right)$. 

\begin{figure}[htb!]
\centering
\includegraphics[scale=0.4]{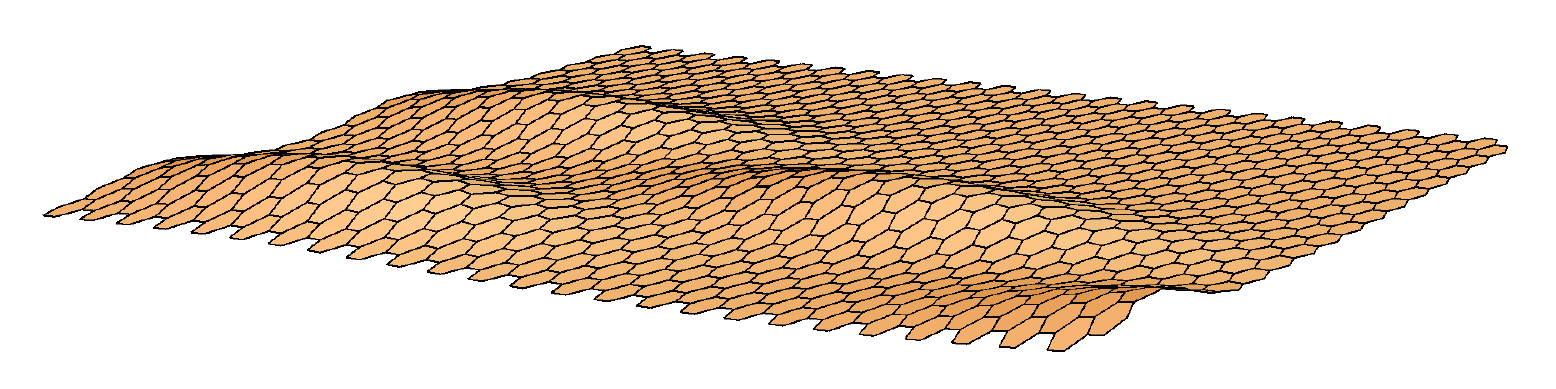}\\
\includegraphics[scale=0.5]{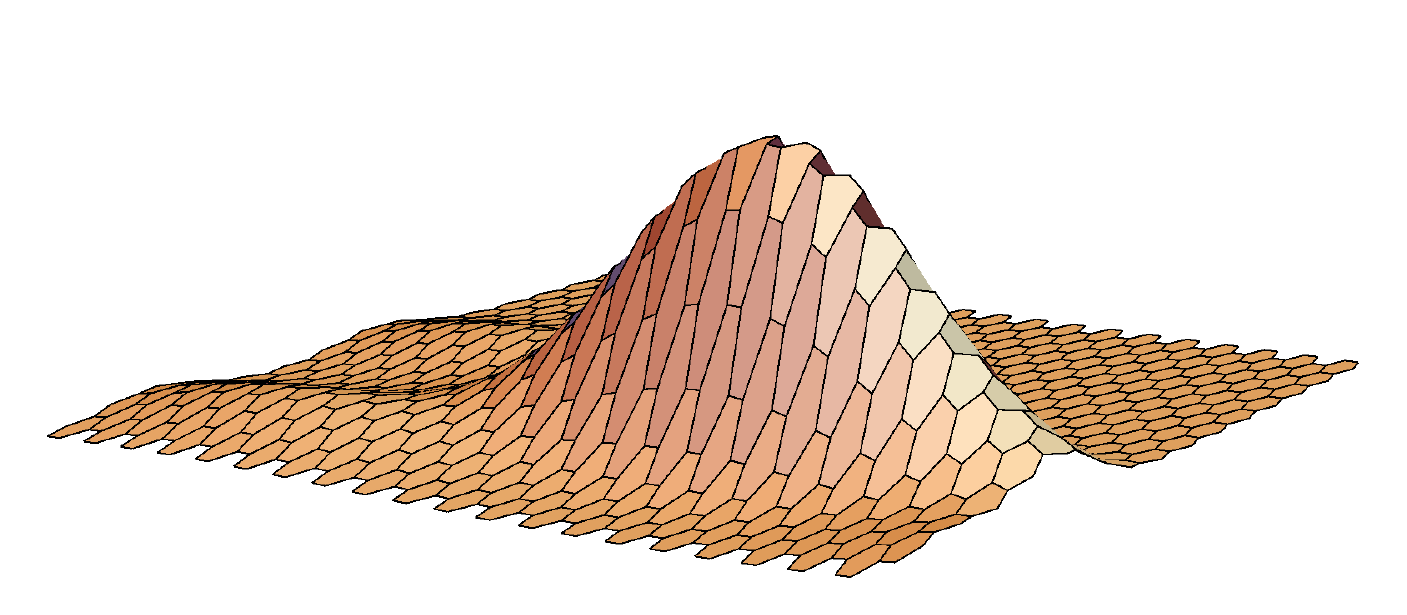}
\caption{\label{figureone} Top and bottom show schematic pictures of graphene in the weak and strong curvature regimes, respectively.}
\label{figesc1}
\end{figure}

The approximations we are going to make here corres\-pond to two different physical situations (see figure \ref{figesc1}). In the first scenario,  the value of the scalar curvature satisfies the condition $RL^2<1$ , where $L$ is a typical graphene sample's size. In this case, one  has an almost flat sheet of graphene with small curvature corrugations. Thus, one can consider $\hat{h}^{2}\equiv \delta^{ij}\hat{p}_{i}\hat{p}_{j}+\hat{h}_{I}$, with $\hat{p}_{j}=-i\partial_{j}$, where $\hat{h}_{I}$ is a perturbation from the flat graphene that  includes all  curvature dependance.  In addition, the coordinates $y$ range over all the domain of the graphene. In the second scenario, the value of the scalar curvature satisfies the condition $1/L^2\ll R<1/d^2$, where $d$ is a typical size of a local domain or patch where graphene presents a pronounced curvature. In this case,  one has a sheet of graphene with strong curvature values at some points where the geometry around these points is considered approximately quadratic, which can be either parabolic or hyperbolic, depending on the sign of the scalar curvature $R$. 
In this case, the coordinates range over all point of the domain of radius $d$.  In what follows, we give expressions for the Green functions in both approximations.
  
\subsection{Small curvature corrugations}
The simplest approximation corresponds with the situation where $\hat{h}_{I}$ consist of a perturbation from the flat ``Hamiltonian" $\delta^{ij}\hat{p}_{i}\hat{p}_{j}$. In this approach,  the Green's function can be written through the resolvent operator $K=1/\left(z^2_{F}-\hat{h}^{2}\right)$ 
which can likewise  be expanded as a perturbation series:
\begin{eqnarray}
K=K_{0}+K_{0}\hat{h}_{I}K_{0}+\cdots, 
\label{resolvent-expansion}
\end{eqnarray}
where the unperturbed resolvent operator is given by $K_{0}\left(p\right)=1/\left(z^2_{F}-\hat{p}^2\right)$.  Next, we need to find an expression for $\hat{h}_{I}$. The first approximation consist of cutting off the series expansion up to linear terms in curvature, then, the pertubation is written as 
\begin{eqnarray}
\hat{h}_{I}\approx \frac{1}{3}R\indices{^{i}_{k\ell}^{j}}\hat{p}_{i}y^{k}y^{\ell}\hat{p}_{j}-i\left(R\indices{^{ab}_{ki}}\delta^{ij}\Sigma_{ab}y^{k}\right)\hat{p}_{j}+\frac{1}{12}R.\nonumber\\
\label{approxIH}
\end{eqnarray}
Next, we sandwich the equation (\ref{resolvent-expansion}) between bra $\left<y\right|$ and ket $\left|0\right>$ and use the resolvent identity $\mathbbm{1}=\int \frac{d^{d}p}{\left(2\pi\right)^{d}}\left|p\right>\left<p\right|$. Also, we simplify the calculation using the relation (\ref{relation}). The interaction contribution of the Green function can be written as 
\begin{eqnarray}
\left<y\left|K_{0}\hat{h}_{I}K_{0}\right|0\right>&=&\int \frac{d^{2}p}{\left(2\pi\right)^2}\frac{d^{2}k}{\left(2\pi\right)^2}
K_{0}\left(p\right)\left<p\left|\hat{h}_{I}\right|k\right>\times\nonumber\\
&&K_{0}\left(k\right)e^{iy\cdot p}\label{Eq15}.\nonumber\\
\end{eqnarray}
Now, it is not difficult to show that the first two terms of $\hat{h}_{I}$  (Eq. (\ref{approxIH})) do not contribute as a consequence of the antisymmetric character of the Riemann curvature tensor (see appendix \ref{apendice}).  Thus, the Green function up to first correction is given by
\begin{eqnarray}
\left(\hbar v_{F}\right)^2G(y,0,z_{F})&=&\frac{1}{\left(g\left(x\right)\right)^{\frac{1}{4}}}\int\frac{d^{2}p}{\left(2\pi\right)^{2}} K_{0}\left(p\right)\left(\mathbbm{1}\right.\nonumber\\&+&\left.\frac{R}{12}K_{0}\left(p\right)\mathbbm{1}+\cdots\right)e^{i y\cdot {\bf p}}.\nonumber\\
\label{expansion2}
\end{eqnarray}
This result for the Green function is the same obtained some time before by L. Parker et al \cite{Parker} in the context of Quantum Field Theory (QFT) in curved space-time. Clearly, one can follow the procedure implemented here, based in the references \cite{Denjoe},  to compute higher order curvature terms. These terms, however, will be left out of the scope of this work.

\subsection{Strong curvature corrugations }

 Within this approximation one retains a  quadratic form of the local geometry. In this sense, we will look for eigenvalues and eigenstates of the operator (\ref{eqexact}).  One can further simplify using $R_{ik\ell j}=\frac{R}{2}\left(\delta_{i\ell}\delta_{kj}-\delta_{ij}\delta_{k\ell}\right)=\frac{R}{2}\epsilon_{ik}\epsilon_{\ell j}$, valid for two-dimensional manifolds, and  $\Omega_{j}=\frac{i}{8}Ry^{i}\epsilon_{ij}\sigma_{3}$ for the spin-connection. Accordingly, the operator is 
\begin{eqnarray}
\hat{h}^{2}&=&\delta^{ij}\left(\hat{p}_{i}+\frac{R}{8}\epsilon_{il}y^{l}\sigma_{3}\right)\left(\hat{p}_{j}+\frac{R}{8}\epsilon_{ik}y^{k}\sigma_{3}\right)\nonumber\\&-&\frac{R}{6}\epsilon_{ik}\epsilon_{\ell j}\hat{p}^{i}y^{k}y^{\ell}\hat{p}^{j}+\frac{R}{12}.
\label{complete-approx}
\end{eqnarray}
Now, using the commutation relation $\left[y^{\ell}, \hat{p}^{i}\right]=i\delta^{\ell i}$, the second term of Eq. (\ref{complete-approx}) can be expressed  as
\begin{eqnarray}
\epsilon_{ik}\epsilon_{\ell j}\hat{p}^{i}y^{k}y^{\ell}\hat{p}^{j}=-\hat{L}^2,
\end{eqnarray}
where $\hat{L}=\epsilon_{ij}y^{i}\hat{p}^{j}$ is the two dimensional angular momentum operator. Further corrections can be included when we consider variations of the curvature.  Clearly, $\hat{h}^2$ can be written as 
\begin{eqnarray}
\hat{h}^{2}=\left(\begin{array}{cc}
\hat{h}^{2}_{+} & 0\\
0 & \hat{h}^{2}_{-}
\end{array}\right),\nonumber
\end{eqnarray}
with 
\begin{eqnarray}
\hat{h}^{2}_{\zeta}=\delta^{ij}{\hat\Pi}_{i}{\hat\Pi}_{j}+\frac{R}{6}\left(\hat{L}^2+\frac{1}{2}\right),
\label{EqOp}
\end{eqnarray}
where ${\hat\Pi}_{i}=\hat{p}_{i}+\zeta_{R}\frac{\left|R\right|}{8}\epsilon_{li}y^{l}$, and $\zeta_{R}=\zeta{\rm sgn}\left(R\right)=\pm 1$. 

Now,  to find the eigenvalues and eigenstates of the operator (\ref{EqOp}) we express the operators $\hat{\Pi}_{i}$ and $\hat{L}$ in terms of creation and annihilation operators. Since we have a two-dimensional problem we introduce the operators $a_{\zeta}$, $a^{\dagger}_{\zeta}$, $b_{\zeta}$ and $b^{\dagger}_{\zeta}$, which are given in terms of the $\hat{X}_{i}$'s,   conjugated operators to the $\hat{\Pi}_{i}$'s, defined by $\hat{X}_{i}=y^{i}-\zeta 4\epsilon^{ij}\hat{\Pi}_{j}/R$.  In appendix \ref{alg-op}, the commutation algebra of the operators $\hat{\Pi}_{i}$ and $\hat{X}_{i}$ is presented. These relations are useful  to show {\small $\left[a_{\zeta}, a^{\dagger}_{\zeta}\right]=\left[b_{\zeta}, b^{\dagger}_{\zeta}\right]=1$} and {\small $\left[a_{\zeta}, b_{\zeta}\right]=\left[a^{\dagger}_{\zeta}, b_{\zeta}\right]=\left[a^{\dagger}_{\zeta}, b^{\dagger}_{\zeta}\right]=\left[a_{\zeta}, b^{\dagger}_{\zeta}\right]=0$}.  The advantage of introducing these operators is that the angular momentum operator can be written as
\begin{eqnarray}
L=-\zeta_{R}\left(b^{\dagger}_{\zeta}b_{\zeta}-a^{\dagger}_{\zeta}a_{\zeta}\right), 
\end{eqnarray} 
and the operator $\hat{h}^{2}_{\zeta}$ as
\begin{eqnarray}
\hat{h}^{2}_{\zeta}=\frac{\left|R\right|}{2}\left(a^{\dagger}_{\zeta}a_{\zeta}+\frac{1}{2}\right)+\frac{R}{6}\left(\hat{L}^2+\frac{1}{2}\right).
\end{eqnarray}
In addition, following up the analogy with the quantum harmonic oscillator one can  diagonalize straightforwardly this operator. The resulting quantum states are 
\begin{eqnarray}
\left| n,m\right>=\frac{1}{\sqrt{\left(n+m\right)!}\sqrt{n!}}\left(b^{\dagger}_{\zeta}\right)^{n+m}\left(a^{\dagger}_{\zeta}\right)^{n}\left|0,0\right>, 
\end{eqnarray}
where the quantum numbers $n$ and $m\equiv \bar{n}-n$, where $\bar{n}$ being the eigenvalue of the number operator $b^{\dagger}_{\zeta}b_{\zeta}$ and $n$ is the eigenvalue of the number operator $a^{\dagger}_{\zeta}a_{\zeta}$. Here, the eigenvalues of the operator $\hat{h}^{2}_{\zeta}$ are denoted by $\lambda^{2}_{n,m}$. These eigenvalues are  $\lambda_{n,m}=\pm E_{n,m}/\hbar v_{F}$, where
\begin{eqnarray}
E_{n,m}=\hbar \omega_{c}\sqrt{n+\frac{1}{2}+\frac{{\rm sgn}\left(R\right)}{3}\left(m^2+\frac{1}{2}\right)}.
\label{energy-level}
\end{eqnarray}
and $\omega_{c}=v_{F}\sqrt{\left|R\right|/2}$ is the cyclotron frecuency in this situation. Notice that $\lambda^{2}_{n,m}$ does not depend on $\zeta$; this means that $\hat{h}^{2}_{+}$ and $\hat{h}^{2}_{-}$ have the same eigenvalues.  This is not a surprise since there is not preference between the triangular lattices that make up the hexagonal lattice. In the above energy eigenvalues  ${\rm sgn}\left(x\right)$ is the sign function.  Also, let us note that for negative values of curvature the quantum number $n$ has a minimum value different from zero, namely, for both sign of curvature by $n_{\rm min}=\left[(m^2+1)/3\right]\Theta\left(-R\right)$.  Additionally, for a finite system of area $A$ the Landau degeneracy, in the symmetric gauge, is given by $m_{\rm max}=\frac{eB_{z}A}{2\pi\hbar}$. In particular, for  $A=\pi d^2$ this means that $m=-m_{\rm max}, \cdots, m_{\rm max}$, where $m_{\rm max}=Rd^2/8$.  In accordance with  our regime of approximation let us notice that the angular momentum modes $m$ are practically not excitated since $Rd^2<1$. Hence,  in this case the eigenvalues modes reduce to $E_{n,0}\equiv E_{n}$. 

Now that we know the quantum states and their associated energy eigenvalues, we may write the spectral decomposition of the Green function
\begin{eqnarray}
G\left(x, x^{\prime},z\right)=\frac{1}{g^{\frac{1}{4}}\left(x\right)}\sum_{m=-m_{\rm max}}^{m_{\rm max}}\sum_{n=n_{{\rm min}}}^{\infty}\frac{\psi^{\dagger}_{n,m}\left(x\right)\psi_{n,m}\left(x^{\prime}\right)}{z^2-E^{2}_{n,m}}\mathbbm{1},\nonumber\\
\label{approx2}
\end{eqnarray}
where $\psi_{n,m}(x)\equiv \left<n,m\right|\left. x\right>$ are normalized eigenfunctions. For the moment, the explicit form of eigenfunctions is not needed in the present work, but they can nonetheless be written in various representations. 

\section{Curvature as pseudo-magnetic field}\label{sec3dot5}

In this section,  we establish an analogy of the Dirac fermions in curved space with the Landau problem.  The squared Hamiltonian of the massless Dirac fermions in a uniform constant field is 
\begin{eqnarray}
\hat{H}^2=v^2_{f}\delta^{ij}\hat{\pi}_{i}\hat{\pi}_{j}+\left(v^2_{F}e\hbar\right)B\sigma_{3},
\label{diracmagnetic1}
\end{eqnarray} 
where the gauge-invariant kinetic momentum is $\pi_{i}=\hat{p}_{i}+\frac{eB}{2}x_{j}\epsilon_{ji}$ and $B$ the uniform magnetic field (see appendix \ref{Landau}).  

Now, in order to establish precisely the analogy between the Landau Hamiltonian for the massless Dirac fermions in curved space we write down again the equation (\ref{EqOp})
\begin{eqnarray}
\hat{h}^{2}_{\zeta}=\delta^{ij}{\hat\Pi}_{i}{\hat\Pi}_{j}+\frac{R}{12}\mathbbm{1}+\frac{R}{6}\hat{L}^2,
\label{EqOpAp1}
\end{eqnarray}
Note that the first term of $\hat{h}_{\zeta}^{2}$ corresponds to a Landau problem Hamiltonian in the symmetric gauge where 
\begin{eqnarray}
 \frac{\hbar R}{4e}\equiv B_{s}
 \label{pseudo-campo}
 \end{eqnarray}
 plays the role of the  ``magnetic field'' or pseudo-magnetic field as it is posed in \cite{guinea}, whereas the second term involves a rotor contribution where the curvature term $R/3$ plays the role of the inverse of the moment of inertia. The cyclotron resonance frecuency satisfies $\hbar \omega_{c}=\sqrt{2e\hbar v^2_{F}B_{s}}$.  These squared Hamiltonians (\ref{diracmagnetic1}) and (\ref{EqOpAp1}) bear two crucial differences, in the magnetic case the curvature strength $B$ is coupled to $\sigma_{3}$ whereas in the curved case the corresponding curvature $R$ is couple to the identity matrix $\mathbbm{1}$. In addition, with the approximation we performed another crucial difference is the $(R/6) \hat{L}^2$ term present in the curved case. Besides these differences one can think of curvature  $R$ as a physical field, analogous to the magnetic field, that is responsible of several phe\-no\-me\-na like the pseudo-Landau states. 

 We emphasize that the relation between curvature $R$ and pseudo-magnetic field can be subjected to experimental scrutiny. In particular, let us focus on figures ${\rm 3(A)}$ and ${\rm 3(B)}$ found on \cite{levy} by  Levy and coworkers. As they mentioned, figure ${\rm 3(A)}$ corresponds to the experimental topographic line scan and the experimental determination of the $B_{s}$ over the tip trajectory shown in black line in their figure ${\rm 3(B)}$. We extracted data of the topographic line scan in order to fit a curve with a fifth order polynomial (see Eq. (\ref{fitcurve}) of appendix \ref{pseudo}) and then computed the curvature $\kappa_{1}$ of this curve at the maximum, this  corresponds to a principal curvature of the graphene nanobubble.  Although we did not access the data of the transverse line, by inspection of the figure 3(B) of the same work \cite{levy} one can state that the radius of curvature $r_{2}$ for a transverse line is bigger than  $r_{1}=1/\kappa_{1}$, which means that the Ricci curvature $R:=2\kappa_{1}\kappa_{2}<2\kappa^2_{1}\simeq 0.58/{\rm nm}^2$.  Now, using this value of curvature one gets the bound  $B_{s}< 600~ {\rm T}$. In others words, the relation between  $R$ and $B_{s}$ looks reasonably in agreement with the experimental observations.

\section{Energy spectrum and density of states of a curved graphene}\label{sec4}

In this section, we give expressions for  the LDOS in both physical scenarios posed in the last section, which are weak curvature corrugation and strong curvature regime on the  graphene sheet. The local density of states (LDOS), $\varrho\left(\epsilon, {\bf x}\right)$,  is used in order to probe the curvature effects on the electronic degrees of freedom of graphene. In particular, LDOS is expressed through the the Green's function \cite{Atland} by
\begin{eqnarray}
\varrho\left(\epsilon, {\bf x}\right)=-\frac{g_{v}}{\pi}{\rm sgn}(\epsilon)\lim_{\delta\to 0}{\rm Im}~{\rm tr}~\mathbb{G}\left({\bf x}, {\bf x}, \epsilon+i\delta\right).
\label{LDOS}
\end{eqnarray}
The $g_{v}=2$ is the valley degeneration due to the two independent Dirac points in graphene and   ${\rm tr}$ refers to  trace over the pseudo-spin and spin space.

\subsection{DOS and energy spectrum of graphene with small curvature corrugations}

In this subsection we compute the local density of states corresponding to curved graphene within the approximation (\ref{expansion2}). Basically, we need to take the pseudo-spin and spin trace, the imaginary part, and the limiting process $y\to 0$ on the function $\mathbb{G}=\left(z+\mathcal{H}_{K}\right)G$. The pseudo-spin trace cancels out the term $\mathcal{H}_{K}G_{F}$ since the matrix structure of $\mathcal{H}_{K}$ is traceless, thus one has ${\rm tr}~\mathbb{G}=4zG$, where $4=2\times 2$ are the pseudo-spin and spin degrees of freedom. To take the imaginary part and the limiting process of $4zG$ one can follow various ways, for instance, one could express $G$ in terms of Bessel functions as done in \cite{Tesis-Paco}. Here, we note that at the level of approximation stated above (\ref{expansion2}), the Green function $G$ is consistent with
\begin{eqnarray}
G\left(y,0,\epsilon_{F}\right)=\frac{g^{-\frac{1}{4}}\left(x\right)}{\left(\hbar v_{F}\right)^2}\int \frac{d^{2}p}{\left(2\pi\right)^{2}}\frac{e^{i {\bf y}\cdot {\bf p}}}{\epsilon^2_{F}-\left(p^2+\frac{R}{12}\right)+i\delta}, \nonumber\\
\end{eqnarray}
where we have shown explicitly the $\delta$ dependance. Now, we take $y\to 0$ or $x\to x^{\prime}$ and recall that $g\left(x^{\prime}\right)=1$,  use  the Dirac identity \cite{diracid} and then we take the imaginary part. Accordingly, the local density of states is given by
\begin{eqnarray}
\varrho\left(\epsilon, {\bf x}\right)=\frac{4g_{v}\left|\epsilon\right|}{\left(\hbar v_{F}\right)^2}\int_{0}^{\infty} \frac{dp}{2\pi}p~\delta\left(p^2-p^2_{0}\right), 
\end{eqnarray}
where $p^2_{0}=\left(\epsilon^2/\left(\hbar v_{F}\right)^2-\frac{R}{12}\right)$. Note that this last relation imposes the condition 
\begin{eqnarray}
\epsilon^2\geq \epsilon^2_{R}\equiv\frac{\left(\hbar v_{F}\right)^2}{4}\frac{\left|R\right|}{3}.
\end{eqnarray}
The Dirac delta can be expanded by means of an identity \cite{diracid} 
\begin{eqnarray}
\delta\left(p^2-p^2_{0}\right)=\frac{1}{2p_{+}}\left[\delta\left(p-p_{+}\right)+\delta\left(p-p_{-}\right)\right],
\end{eqnarray}
where $p_{\pm}=\pm p_{0}$. Then, the LDOS is given by
\begin{eqnarray}
\varrho\left(\epsilon, {\bf x}\right)= \frac{2\left|\epsilon\right|}{\pi \left(\hbar v_{F}\right)^2}\Theta\left(\epsilon^2- {\rm sgn}\left(R\right)\epsilon^2_{R}\right)
\label{ldos}
\end{eqnarray}
where $\Theta(x)$ is the Heaviside function.  Clearly, for flat spaces $\epsilon_{R}=0$,  we recover the standard relation for the local density of states \cite{Katnelson}. Also, it is noted that, within this approximation,  for points ${\bf x}$ with negative va\-lues of curvature the local density of states corresponds to the usual one.  However, for points ${\bf x}$ with positive values of curvature we find a change in the local density of states.  These results for the LDOS can be improved with higher order curvature corrections of the Green function \cite{Castro-future1}. In addition, this same result (Eq. (\ref{ldos})) can be seen at the level of the dispersion relation $E^2(p):=\int \frac{d^{2}q}{\left(2\pi\right)^2}\left<p\left|g\mathcal{H}^2_{K}\right|q\right>$, which can be computed using the operator $\hat{h}^{2}$. The energy dispersion relation turns out to be 
\begin{eqnarray}
E(p)=\pm v_{F}\sqrt{p^2+\frac{\hbar^2 R}{12}}.
\label{disprel}
\end{eqnarray}  
Note that this energy dispersion relation is similar to the relativistic energy dispersion relation where $\hbar\sqrt{R/12}$ plays  the role of mass. The curvature appears with a prefactor $\hbar^2$ which makes $v^2_{F}\hbar^2 R/12<1.42~{\rm eV}^2$ for a sample of size $L\sim 10~{\rm nm}$.  In the positive curvature case, we have shown that in graphene arise  a small gap,  whereas in the negative curvature case we found the manner in which the curvature induces a minimum value of the charge carriers momentum. These both cases modify the conical nature of the energy dispersion relation  turning  on a shift in the momentum space of the Dirac cone.

\subsubsection{Example: Curvature corrections on the internal energy.} In order to look for a more tangible curvature effect, let us evaluate the electronic internal energy, $U$, for a sheet of graphene with  weak curvature. Taking into account particles and holes, $U$ can be computed using the following expression:
 \begin{eqnarray}
U=2\int_{0}^{\infty} d\epsilon \rho\left(\epsilon\right) \frac{\epsilon}{\exp\left(\beta \epsilon\right)+1},
\label{internal}
\end{eqnarray}
where $\beta=1/k_{B}T$  and $\rho\left(\epsilon\right)=\int dA~ \varrho\left(\epsilon,{\bf x}\right)$ is the density of states (DOS), with  $dA$ being the area element of the surface. Now, we substitute the LDOS (\ref{ldos}) and we exchange integrals, to get
\begin{eqnarray}
U\simeq\frac{2}{\pi \left(\hbar v_{F}\right)^2\beta^{3}}\int dA \int^{\infty}_{\beta^2\Theta\left(R\right)\epsilon^2_{R}}\frac{dz\sqrt{z}}{\exp\left(\sqrt{z}\right)+1}.
\end{eqnarray}
For $\epsilon_{R}=0$, one can integrate and find that the internal energy per unit of area, $A$, is given by the standard result \cite{diracmaterials}
\begin{eqnarray}
u=\frac{6\zeta\left(3\right)}{\pi}\left(k_{B}T\right)^{3}/\left(\hbar v_{F}\right)^{2}.
\label{int-energ}
\end{eqnarray}
For curved graphene, $\epsilon_{R}\neq 0$, we performed a perturbative expansion in powers of $\beta \epsilon_{R}$ and computed the heat capacity per unit of area, $c\left(T\right)$ 
\begin{eqnarray}
c(T)/k_{B}&=&\frac{18\zeta\left(3\right)}{\pi}\left(\frac{k_{B}T}{\hbar v_{F}}\right)^2\nonumber\\
&-&\frac{1}{144\pi}\left[\frac{1}{A}\int_{\mathcal{D}} dA ~
R^{2}\right]\left(\frac{\hbar v_{F}}{k_{B}T}\right)^2+\cdots,\nonumber\\
\end{eqnarray}
where $\mathcal{D}$ is a local domain with positive curvature. Note that, at the level of this approximation, negative  values of curvature do not make any correction to the heat capacity of the graphene.

\subsection{DOS and energy spectrum of graphene with strongly curved protuberance}  

In this section, we give a formal expression for the density of states in terms of the energy eigenvalues using the approximation (\ref{approx2}). It is assumed, for simplicity, that curvature is approximately constant. Similarly to the weak curvature regime  one has ${\rm tr} \mathbb{G}=4z G$. Next, we use the orthogonality property of $\psi_{n,m}\left(x\right)$, the limit $x\to x^{\prime}$ and the Dirac identity \cite{diracid}. Finally, we use the fact that within the strong curvature approximation the angular momentum is not excitable. In this manner, the density of states is given by
\begin{eqnarray}
\rho\left(\epsilon\right)\simeq 8g_{v}m_{\rm max}\left|\epsilon\right|\sum_{n=0}^{\infty}\delta\left(\epsilon^2-E^2_{n}\right). 
\label{ldos2}
\end{eqnarray}
The energy eigenvalues  of the Dirac fermions moving in a positive and negative local curved patch of a manifold are given by 
\begin{eqnarray}
E_{n}=\hbar\omega_{c}\left\{\begin{array}{c}\sqrt{n+\frac{2}{3}}~~~{\rm R}>0,\\
\sqrt{n+\frac{1}{3}}~~~{\rm R}<0,
\end{array}\right.
\end{eqnarray}
where we recall the resonance cyclotron frecuency $\hbar \omega_{c}=\hbar v_{F}\sqrt{\left|R\right|/2}=\sqrt{2e\hbar v^2_{F}B_{s}}$. These eigenvalues spectrum is almost the same as the one obtained for flat graphene in presence of a strong and transversal magnetic field \cite{Katnelson}. 

\subsubsection{Example: Curvature corrections on the internal energy in the strong curvature regime.} 
In what follows, we give the expression for  the internal energy for the strong curvature regime. We substitute the DOS in (\ref{internal}) and we use the identity \cite{diracid}. After a straightforward calculation, the internal energy and the number of particles (particles and holes) are 
\begin{eqnarray}
U\left(T\right)&\simeq&8g_{v}m_{\rm max}\sum_{n=0}^{\infty}E_{n}F\left(\beta \left(E_{n}-\mu\right)\right),\nonumber\\
N&\simeq&8g_{v}m_{\rm max}\sum_{n=0}^{\infty}F\left(\beta \left(E_{n}-\mu\right)\right),
\end{eqnarray}
where $F(x)=1/(\exp(x)+1)$ is the Fermi-Dirac statistics. Also, we have introduced a chemical potential $\mu$. The energy eigenvalues are proportional to $\hbar\omega_{c}$, thus it is natural to define a characteristic temperature by $k_{B}T_{R}=\hbar v_{F}\sqrt{\left|R\right|/2}=\hbar\omega_{c}$. 

As a matter of consistency, one must recover the standard internal energy per unit of area (\ref{int-energ}) in the limit when $T\gg T_{R}$. Indeed, one can make the steps in the appendix \ref{cp} to reach this statement.   Now, the opposite limit $T\ll T_{R}$ is usually the case when one has a pronounced value of curvature. For instance, the pseudo-magnetic field, $B_{s}$, ranges the values $\left[300, 400\right]{\rm T}$ in the Levy et al. experiment \cite{levy}. These values correspond to the cha\-rac\-te\-ris\-tic temperature $T_{R}$ in the range $\left[18, 21\right]\times 10^{3}{\rm K}$. Hence, the thermal effects are practically negligible for massless Dirac fermions confined in an strong curvature protuberance. Thus, the internal energy corresponds essentially to the energy of the ground state. 

The energy of the ground state can be computed assumed that  all pseudo-Landau levels are filled up to the Fermi level $n_{F}+1=\frac{N}{8g_{v}m_{{\rm max}}}=\frac{\hbar\pi \sigma}{4g_{v}e B_{s}}$, where $\sigma=N/A$ is the density of the particles in the domain where the curvature is pronounced.  Thus the energy of the ground states, $U\left(0\right)\equiv \lim_{T\to 0} U\left(T\right)$, is 
\begin{eqnarray}
\frac{U\left(0\right)}{A}=\frac{8e}{\pi \hbar}B_{s}\sum_{n=0}^{\left[\frac{\hbar\pi \sigma}{8e B_{s}}\right]}E_{n}
\end{eqnarray}
This energy will have discontinuities in the pseudo-magnetization $\mathcal{M}=-\left({\partial U/\partial B_{s}}\right)_{\sigma}$ at the pseudo-magnetic field values 
\begin{eqnarray}
\left(B_{s}\right)_{n}=\frac{\hbar\pi \sigma}{4eg_{v}\left(1+n\right)}.
\end{eqnarray}
We expect, then, an analogous de Haas-van Alphen effect occurs in this case. In a future communication we will report a systematic study of this analogous phenomena.

\section{Concluding remarks }\label{sec5}

In this paper, we have studied the electronic degrees of freedom of a graphene curved sheet upon the basis of the Dirac equation on curved space-time. The curved space-time we use corresponds to the metric $ds^2=-v^2_{F}dt^2+g_{ij}dx^{i}dx^{j}$, where $v_{F}$ is the Fermi velocity and $g_{ij}$ are the components of the metric tensor of the spatial sector. These electronic degrees of freedom are studied by means of the Local Density of States (LDOS), which is expressed through the Green function associated to the operator $\epsilon-\mathcal{H}_{K}$, where $\epsilon$ is the energy and $\mathcal{H}_{K}$ is an Euclidean Dirac operator. It was studied in two different physical situations: I) weak curvature regime and II) strong curvature regime.

In the first case I), we found that LDOS has an open gap proportional to $\sqrt{R}$ for positive values of curvature whereas LDOS corresponds to the flat expression for the negative values of curvature. In the second case II), it was found an analogy with the well known Landau problem of the flat graphene under a strong magnetic field. The pronounced curvature, in this situation, affects the charge carriers in such a manner that pseudo Landau states emerge. This analogy allows us to identify the curvature with a pseudo-magnetic field $B_{s}$ through the relation, $B_{s}=\hbar R/4e$.  The resonance cyclotron frecuency turns out to be identical to the magnetic case  $\hbar \omega_{c}=\sqrt{2e\hbar v^2_{F} B_{s}}$. The particular differences with the original Landau problem were briefly discussed.

In addition, using the values of the pseudo-magnetic field $B_{s}\sim 400~{\rm T}$ reported in \cite{levy} it was found that the characteristic temperatures are around $T_{R}\equiv \hbar \omega_{c}/k_{B}\sim 2.1\times 10^{4}~{\rm K}$. This means that in the strong curvature case the thermal effects are practically negligible and the internal energy is essentially the energy of the ground state.  In particular, we give an expression for the ener\-gy of the ground state and we remarked the expectation of an analogous de Haas-van Alphen effect.  In a future communication we will report a systematic study of this analogous phenomena.


Our approach can be extended in various directions. As was mentioned before, the calculations performed in the weak curvature regime can be improved including higher order curvature terms in the Green function of the Dirac equation. In the strong curvature regime, it requires a systematic analysis of the oscilations in the pseudo-magnetic field that appear in the de Haas-van Alphen effect; one can take a guide study of the magnetic case (see for instance \cite{Zhang}). In addition, the pseudo-Landau states found here can also be precursive to other analogous effects like the quantum Hall effect and the Khalatnikov-de Haas van Alphen effect. 

\acknowledgments

The author PCV would like to thank Idrish Huet Hern\'andez, Olindo Corradini and Oscar V\'azquez Rodr\'iguez for many valuable discussions. Financial support by Conacyt (Grant Nos. 237425, 289334, and Red Tem\'atica de la Materia Condensada Blanda) and PROFOCIE/2015  is acknowledged.

\appendix
\section{Some extra calculations}\label{apendice}
\subsection{Computation of $\left<y\left|K_{0}\hat{h}_{I}K_{0}\right|0\right>$}

Our starting point to compute this amplitude is the Eq. (\ref{Eq15}). The interaction operator $\hat{h}_{I}$ can be split into three terms
\begin{eqnarray}
\hat{h}^{\left(1\right)}_{I}&=&\frac{1}{3}R\indices{^{i}_{kl}^{j}}\hat{p}_{i}y^{k}y^{l}\hat{p}_{j}\nonumber,\\
\hat{h}^{\left(2\right)}_{I}&=&-iR\indices{^{ab}_{ki}}\delta^{ij}\Sigma_{ab}y^{k}\hat{p}_{j}\nonumber,\\
\hat{h}^{\left(3\right)}_{I}&=&\frac{1}{12}R\nonumber.
\end{eqnarray}
The amplitude associated to the last term is straightforward. For the first and second operator it is convenient to use the relation 
\begin{eqnarray}
\left<p\left|y^{I}\right|k\right>=\left(2\pi\right)^{2}i^{m}\frac{\partial^{m}}{\partial k^{I}}\delta\left({\bf p}-{\bf k}\right),
\label{relation}
\end{eqnarray}
where $I$ is a collective indice of order $m$ which means that $y^{I}=y^{i_{1}}y^{i_{2}}\cdots y^{i_{m}}$. For the first operator, $\hat{h}^{\left(1\right)}_{I}$, one has to use $\left<p\left|y^{k}y^{l}\right|k\right>$. After substituting in the amplitude one has
\begin{eqnarray}
\left<y\left|K_{0}\hat{h}^{\left(1\right)}_{I}K_{0}\right|0\right>&=&-\frac{1}{3}R\indices{^{iklj}}\int \frac{d^{2}p}{\left(2\pi\right)^{2}}K_{0}\left(p\right)\times\nonumber\\&&p_{i}\partial_{k}\partial_{l}\left(p_{j}K_{0}\left(p\right)\right)e^{ip\cdot y}\nonumber,\\
\end{eqnarray}
where $\partial_k=\frac{\partial}{\partial p^{k}}$. Next, one can show that 
\begin{eqnarray}
p_{i}\partial_{k}\partial_{l}\left(p_{j}K_{0}\left(p\right)\right)&=&K_{1}(p)\left[\delta_{lj}p_{i}p_{k}+\delta_{kj}p_{i}p_{l}+\delta_{kl}p_{i}p_{j}\right]\nonumber\\&+&K_{2}(p)p_{i}p_{j}p_{k}p_{l}, \nonumber\\
\end{eqnarray}
where $K_{1}=K^{\prime}_{0}/p$ and $K_{2}=K^{\prime}_{1}/p$. Thus one has a symmetric tensor in the indices $l$ and $j$, then by contracting with the Riemman tensor, which is antisymmetric in these indices, one gets that the amplitude is exactly cero. 

In a similar manner, for the second term one gets
\begin{eqnarray}
\left<y\left|K_{0}\hat{h}^{\left(1\right)}_{I}K_{0}\right|0\right>&=&-R^{abkj}\Sigma_{ab}\int \frac{d^{2}p}{\left(2\pi\right)^{2}}K_{0}\left(p\right)\times\nonumber\\
&&\partial_{k}\left(p_{j}K_{0}\left(p\right)\right)e^{ip\cdot y}.\nonumber\\
\end{eqnarray}
This amplitude is also cero since $\partial_{k}\left(p_{j}K_{0}\left(p\right)\right)=\delta_{kj}K_{0}+p_{j}p_{k}K_{1}$ is symmetric in the indices $k$ and $j$. 

\subsection{Algebra of operators}\label{alg-op}

The creation and anihilation operators are defined by 
\begin{eqnarray}
a_{\zeta}&=&\sqrt{\frac{2}{\left|R\right|}}\left(\hat{\Pi}_{x}-i\zeta_{R}\hat{\Pi}_{y}\right),\nonumber\\
a^{\dagger}_{\zeta}&=&\sqrt{\frac{2}{\left|R\right|}}\left(\hat{\Pi}_{x}+i\zeta_{R}\hat{\Pi}_{y}\right),
\end{eqnarray}
and
\begin{eqnarray}
b_{\zeta}&=&\sqrt{\frac{\left|R\right|}{8}}\left(\hat{X}+i\zeta_{R} \hat{Y}\right)\nonumber\\
b^{\dagger}_{\zeta}&=&\sqrt{\frac{\left|R\right|}{8}}\left(\hat{X}-i\zeta_{R} \hat{Y}\right),
\end{eqnarray}
where $\hat{X}_{i}$'s are conjugated operators to the $\hat{\Pi}_{i}$'s defined by $\hat{X}_{i}=y^{i}-\zeta 4\epsilon^{ij}\hat{\Pi}_{j}/R$.

 The set of operators $\{\hat{\Pi}_{i}, \hat{ L}\}$ form a closed Lie algebra. In particular, they satisfy the commutation relations 
 \begin{eqnarray}
 \left[\hat\Pi_{i}, \hat\Pi_{j}\right]&=&-i\zeta R\epsilon_{ij}/4\nonumber,\\
  \left[\hat{L},\hat\Pi_{i}\right]&=&i\epsilon_{ik}\hat\Pi^{k},\nonumber\\
  \left[\hat{L}, \hat\Pi^2_{i}\right]&=&0.
  \end{eqnarray}
Then $\hat{\Pi}^2_{i}$ and $\hat{\bf L}$ can be diagonalize simultaneously.  Alternatively, one can define the operators $\hat{X}_{i}=y^{i}-\zeta \frac{4}{R}\epsilon^{ij}\hat{\Pi}_{j}$, whose commutation relations are 
\begin{eqnarray}
{\small\left[\hat{X}_{i},\hat{X}_{j}\right]=4\zeta i\epsilon_{ij}/R},
\end{eqnarray}
which commute with each $\hat{\Pi}_{i}$; in particular, the angular momentum operator can be written as 
\begin{eqnarray}
L=\frac{2\zeta}{R}\delta^{ij}{\hat\Pi}_{i}{\hat\Pi}_{j}-\frac{\zeta R}{8}\delta^{ij}{\hat {X}}_{i}{\hat {X}}_{j}.\nonumber
\end{eqnarray}

\subsection{Landau Hamiltonian for massless Dirac fermions }\label{Landau}

In this section we can follow \cite{Katnelson} for the treatment of massless Dirac fermions in presence of a uniform magnetic field. Let us begin with the Landau Hamiltonian in this context
\begin{eqnarray}
\hat{H}=v_{F}\sigma\cdot\left(\hat{\bf p}+e{\bf A}\left({\bf x}\right)\right),
\end{eqnarray}
where ${\bf A}\left({\bf x}\right)$ is the vector potential. To obtain the eigenvalues and eigenstates of $\hat{H}$, it is often common to square the operator. Thus, after a straigforward calculation it is not difficult to find
\begin{eqnarray}
\hat{H}^2=v^2_{F}\delta_{ij}\hat{\pi}_{i}\hat{\pi}_{j}+\frac{1}{2}v^2_{F}e\hbar \epsilon_{ij}F_{ij} \sigma_{3},
\label{square}
\end{eqnarray}
where the gauge-invariant kinetic momentum $\pi_{i}=\hat{p}_{i}+\frac{eB}{2}x_{j}\epsilon_{ji}$ and the curvature strength tensor $F_{ij}=\partial_{i}A_{j}-\partial_{j}A_{i}$. The squared Hamiltonian, (\ref{square}), is in fact, a Lichnnerowiczs formula for the Dirac operator on a manifold. In addition, for an uniform magnetic field one has the vector potential in the symmetric gauge : $A_{i}=\frac{B}{2}x_{j}\epsilon_{ji}$, then $\epsilon_{ij}F_{ij}=2B$. Which yields the following squared Hamiltonian
\begin{eqnarray}
\hat{H}^2=v^2_{F}\delta_{ij}\hat{\pi}_{i}\hat{\pi}_{j}+\left(v^2_{F}e\hbar\right) B \sigma_{3}.
\label{diracmagnetic}
\end{eqnarray}

\subsection{Estimation of pseudo-magnetic field} 
\label{pseudo}

In this subsection, we estimate the value of the pseudo-magnetic field based on the extracted points of figure 3(A) of the article by Levy and coworkers. This data corresponds to a section of the actual surface that represents the graphene nanobubble.  Once we have extracted the data, we fit a curve with a simple polynomial approximation. The best fit found corresponds to a fifth order polynomial
\begin{eqnarray}
z\left(x\right)=\sum_{k=0}^{5}\beta_{k}x^{k},
\label{fitcurve}
\end{eqnarray}
where values $\beta_{k}$, with $k=0,\cdots, 5$ are given by $\beta_{0}=-0.000759452$, $\beta_{1}=0.287848$, $\beta_{2}=-0.232805$, $\beta_{3}=0.339021$, $\beta_{4}=-0.180358$, and $\beta_{5}=0.0281599$. This function allows us to compute the curvature of the curve by using the formula
\begin{eqnarray}
\kappa_{1}\left(x\right)=\frac{z^{\prime\prime}\left(x\right)}{\left(1+\left(z^{\prime}\left(x\right)\right)^2\right)^{\frac{3}{2}}}.
\end{eqnarray}
This curvature corresponds to a principal curvature of the surface along the topographical line measured in \cite{levy}. Unfortunately, we do not know the values of the transversal curve so as to compute the value of the other principal curvature. However, glimpsing the figure 3(B) of \cite{levy} one can say that the radius of curvature of the transverse line is bigger than the actual line, i.e. t $r_{2}>r_{1}$, so $\kappa_{2}<\kappa_{1}$, and the Ricci curvature is $R=2\kappa_{1}\kappa_{2}<2\kappa^2_{1}(x_{\rm max})=0.58/{\rm nm}^2$. Using now the Eq. (\ref{pseudo-campo}) one gets the corresponding bound on the pseudo-magnetic field $B_{s}<600~{\rm T}$, which looks reasonably in agreement with experimental observations. 

\subsection{Consistency proof}\label{cp}

We substitute the DOS in (\ref{internal}) and use the identity \cite{diracid}. After a strightforward calculation, the internal energy  (particles and holes) is 
\begin{eqnarray}
U\left(T\right)&=&4g_{v}\sum_{m=-m_{\rm max}}^{m_{\rm max}}\sum_{n=n_{\rm min}}^{\infty}E_{n,m}F\left(\beta \left(E_{n,m}-\mu\right)\right),\nonumber\\
\end{eqnarray}
where we have taken into account the angular momentum modes.  Let us take positive curvature, $R>0$. In addition, define $q=\left(\hbar v_{F}\right)^{2}\frac{R}{2} n$ such that the energy levels (\ref{energy-level}) are given by 
\begin{eqnarray}
E_{q,m}=\sqrt{q+\left(\hbar v_{F}\right)^{2}\frac{R}{6}\left(m^{2}+2\right)}
\end{eqnarray}
Notice that   for each $m$ value  $E_{q, m}$ is almost a continuos spectra for $\left(\hbar v_{F}\right)^{2}\frac{R}{2}$ small. This means that we can replace $q$  by a conti\-nuos variable.  Thus, within this appro\-xi\-ma\-tion the internal energy is given by
\begin{eqnarray}
U\left(T\right)\approx\frac{8g_{v}}{\left(\hbar v_{F}\right)^{2}R}\sum_{m=-m_{\rm max}}^{m_{\rm max}}\int_{0}^{\infty}dq\frac{E_{q,m}}{\exp\left(\beta E_{q,m}\right)+1}\nonumber\\
\end{eqnarray}
We perform the change of variable $p=E_{q,m}$, hence
\begin{eqnarray}
U\left(T\right)\approx\frac{8g_{v}}{\left(\hbar v_{F}\right)^{2}R}\sum_{m=-m_{\rm max}}^{m_{\rm max}}\int_{E_{0, m}}^{\infty}\left[\frac{dp~p^{2}}{\exp{\left(\beta p\right)}+1}\right].\nonumber\\
\end{eqnarray}
Now, using $m_{\rm max}=RL^2/8$, due to Landau degeneracy, and the area $A=\pi L^2$, it is not difficult to conclude that the internal energy per unit area is given by the desired expression (\ref{int-energ}).


\begin{thebibliography}{10}
\bibitem{diracid} The Dirac identity is given by $\frac{1}{x+i\delta}=-i\pi \delta\left(x\right)+{\rm P}\left(\frac{1}{x}\right)$, where $P$ means the principal value. In addition the Dirac delta has the following identity
$\delta\left(x^2-a^2\right)=\left[\delta\left(x-\left|a\right|\right)+\delta\left(x+\left|a\right|\right)\right]/(2\left|a\right|)$.
\end{thebibliography}
\end{document}